\documentstyle[aps,preprint,epsf]{revtex}

\begin{document}

\draft
 
\title{Driving, conservation and absorbing states in sandpiles}
\author{Alessandro Vespignani$^{1}$,
Ronald Dickman$^{2}$,
Miguel A. Mu\~noz$^{1,3}$, and 
Stefano Zapperi$^{4}$}

\address{
$^1$ The Abdus Salam International Centre for Theoretical Physics (ICTP) 
P.O. Box 586, 34100 Trieste, Italy\\
$^2$ Departamento de F\'{\i}sica,
Universidade Federal de Santa Catarina,
Campus Universit\'ario\\ 
Trindade, CEP 88040-900, 
Florian\'opolis --- SC, Brazil\\
$^3$ Dipartimento di Fisica e unit\`a INFM,
Universit\'a di Roma `` La Sapienza", Piazzale A. Moro 2,
I-00185 Roma, Italy\\
$^4$PMMH-ESPCI,
10, rue Vauquelin, 75231 Paris CEDEX 05, France \\
}
 
\date{\today}

\maketitle
\begin{abstract}

We use a phenomenological field theory, reflecting the symmetries
and conservation laws of sandpiles, to
compare the driven dissipative sandpile,
widely studied in the context of self-organized criticality,
with the corresponding fixed-energy model.  The latter
displays an absorbing-state phase transition
with upper critical dimension $d_c=4$. 
We show that the driven model
exhibits a fundamentally different approach to the critical point, 
and compute a subset of critical exponents.
We present numerical simulations in support of our 
theoretical predictions.
\end{abstract}

\pacs{PACS numbers: 64.60.Lx, 05.40.+j, 05.70.Ln }

A wide variety of nonequilibrium systems display transitions
between ``active'' and ``absorbing'' states:
examples are epidemic processes \cite{epi} catalysis
\cite{RDPRV}, directed percolation (DP)\cite{kinzel},
and the depinning of interfaces in quenched disorder \cite{barabasi}.
When driven continuously, such systems may exhibit       
stick-slip instabilities, or
broadly distributed avalanches, commonly associated with 
self-organized criticality (SOC)\cite{btw,grin}. 

SOC sandpiles \cite{btw} possess an infinite number of absorbing
configurations (i.e., from which the system cannot escape),
and are placed, by definition, at the critical point in a two-dimensional
parameter space \cite{vz,dvz} resembling that of directed percolation (DP)
\cite{kinzel} or contact processes \cite{cp,pcp,MUNOZ}.

Under an external drive (i.e., input of particles
at rate $h$), the system jumps among
absorbing configurations via avalanche-like
rearrangements.
Close to the absorbing-state phase transition, a slow drive
induces avalanches whose size distribution decays 
as a power law --- the hallmark of SOC. 
What distinguishes the sandpile from other models with absorbing states
is a {\it conservation law}: avalanche dynamics
conserves the number of grains of ``sand,'' and  
the order parameter is coupled to this conserved field \cite{dvz}.

In this Letter we use a phenomenological 
field theory of sandpiles to show 
how conservation alters the phase transition.
The critical behavior
for $h \to 0$ (the SOC limit)
{\it differs} from that for $h \equiv 0$.
In particular,   
when driving and dissipation are absent, 
the sandpile shows an absorbing-state 
phase transition (with $d_c=4$).
Our approach clarifies the effect of driving
on dynamic phase transitions, and resolves several long-standing
issues regarding sandpiles, such as the upper critical
dimension, the effect of conservation on critical 
exponents, and universality classes \cite{tb,sornette,chessa,lubeck,diaz}.
We perform extensive simulations to 
check our theoretical predictions.
  
In sandpiles \cite{btw},
each site $i$ of a $d-$dimensional lattice bears an integer variable 
$z_i \geq 0$, which we call {\em energy}. 
When a site reaches or exceeds a threshold
$z_c$ it topples: $z_i \to z_i-z_c$, and $z_j \to z_j+1$ 
at each of the $g$ nearest neighbors of $i$. Energy is fed into the
system at rate $h$, and is dissipated at rate $\epsilon$ 
during toppling \cite{vz}. 
At each time step, each site has a probability 
$\propto h$ to receive an energy grain; in each toppling, a grain
is lost with probability $\propto \epsilon$. These rules generalize
the original Bak, Tang and Wiesenfeld (BTW) sandpile automaton \cite{btw}, 
which is recovered in the limit
$h\to 0$ and $\epsilon \to 0$ 
\cite{vz,dvz}.  While the BTW model restricts dissipation to
the boundaries, we focus on (conceptually simpler)  
bulk dissipation; most conclusions apply
to the boundary dissipation case as well.
We also consider the Manna sandpile,
in which $z_c = 2$ and two neighboring sites are chosen at random to
receive energy\cite{manna}.

In the slow-driving limit, each energy addition is followed by an
avalanche of $s$ topplings; 
the avalanche distribution has the scaling form 
$P(s)=s^{-\tau}G(s/s_c)$, where the cutoff scales as  $s_c\sim\xi^D$. 
The correlation length $\xi$ scales with dissipation 
as $\xi\sim\epsilon^{-\nu}$, and is related to the characteristic
avalanche duration $t_c\sim\xi^{z}$.

The order parameter is $\rho_a$,
the density of active sites (i.e., whose height $z \geq z_c$) 
\cite{vz,dvz}; if $\rho_a=0$ the system has reached 
an absorbing configuration. In a coarse-grained description, 
we study the dynamics of a local order-parameter field
$\rho_a({\bf x},t)$,
bearing in mind that
the energy density $\zeta({\bf x},t)$ is
(for $\epsilon=h=0$), a conserved field.
Variations of the 
local energy density are due to:
(i) the external field, $h$;
(ii) dissipation attending toppling: 
$-\epsilon \rho_a$;
(iii) a diffusion-like contribution:
$(1-\epsilon)\nabla^2 \rho_a$.
The latter arises because a gradient in $\rho_a$
leads to a current: the excess in the mean number
of particles arriving at ${\bf x}$ from the left,
over those arriving from the right, 
is $j_x ({\bf x},t) = -(1-\epsilon)\partial_x \rho_a$.  
The net inflow of particles at {\bf x} is 
therefore $-\nabla \cdot {\bf j} = (1-\epsilon)\nabla^2 \rho_a$.
Defining an energy diffusion constant $D_\zeta \propto 1-\epsilon$,
we write the continuity equation for the energy density,
\begin{equation}
\frac{\partial \zeta ({\bf x},t)}{\partial t}=
D_\zeta\nabla^2 \rho_a ({\bf x},t)
-\epsilon \rho_a ({\bf x},t)  + h ({\bf x},t) +\eta_\zeta({\bf x},t),
\label{cons}
\end{equation}
where the driving field  $h({\bf x},t)=
\overline{h} +\eta_h({\bf x},t)$,
with $\overline{h}$ a nonfluctuating term and  $\eta_h$ zero-mean, 
uncorrelated Gaussian noise.
The last term is
dynamically generated {\em Reggeon-field-theory-like} (RFT) noise
\cite{RFT}
$\eta_\zeta({\bf x},t)\sim\sqrt{ \rho_a({\bf x},t)}\eta({\bf x},t)$,
with $\eta$ uncorrelated Gaussian noise. This term vanishes,
as it must, in the absorbing state, $\rho_a = 0$.

The equation for the order-parameter
field is readily obtained \cite{dvz} by extending
the mean-field theory (MFT) of Ref.~\cite{vz}. 
With  $\rho_c({\bf x},t)$ the local density
of ``critical'' sites (i.e., with height $z_c-1$), we have
$$\frac{\partial \rho_a({\bf x},t) }{\partial t} =
D_a\nabla^2 \rho_a({\bf x},t) - \rho_a({\bf x},t) $$
\begin{equation}
+(g-\epsilon)\rho_a({\bf x},t) \rho_c({\bf x},t)  + h \rho_c({\bf x},t)  
+ \eta_a({\bf x},t),
\label{active}
\end{equation}
where 
$\eta_a({\bf x},t)$ is a RFT-like noise whose amplitude is 
proportional to $\sqrt{\rho_a({\bf x},t)}$.
The first two terms represent
toppling\cite{top}; the terms $\propto \rho_c$ represent
critical sites becoming active upon receiving energy, whether from
the external drive, or from toppling neighbors.

In the stationary state, we can avoid ensnarement in an infinite hierarchy
of equations \cite{dvz}, by eliminating $\rho_c$ in favor of $\zeta$ and
$\rho_a$.  In the Manna model, we simply invoke normalization:
$\rho_c = \zeta - z_a \rho_a$,
where $z_a \geq 2 $ is the mean height of active sites.
For BTW we use the phenomenological {\em Ansatz}: 
$\rho_c ({\bf x},t)=[1-\rho_a ({\bf x},t)]f[\zeta({\bf x},t)]$.
That is, the fraction $f$ of nonactive sites that are critical
can be expressed as a single-valued function of the
local energy density. In the slowly-driven
stationary state,  
$\zeta \simeq \zeta_s$ and $f \simeq \rho_c^\infty$, 
where $\zeta_s$ and $\rho_c^\infty$ are
the stationary average values of the energy and the
critical-site density.
We expand $f(\zeta) = \rho^\infty_c + A\Delta \zeta  
+ \cdots$, where $\Delta \zeta \equiv \zeta({\bf x},t) -\zeta_s$,
and $A>0$. We test the validity of these assumptions
by simulating the two-dimensional BTW model on a
lattice of $80 \times 80$ sites, at $\zeta =
\zeta_s = 2.125$.   
To determine $f(\zeta)$, we measure the average energy,
and active- and critical-site densities
in cells of $10 \times 10$ sites.
The conditional probability
density $P(f|\zeta)$ is unimodal and peaked (see Fig.1);
the mean increases linearly with $\zeta$, indicating
that $\rho_c$ is well-approximated by $(1-\rho_a)f(\zeta)$,
with $f$ linear in the neighborhood of $\zeta_s$.
     
Eqs.~(\ref{cons}) and (\ref{active}), describing the coarse-grained
dynamics
of sandpiles, resemble the field-theory
for the pair contact process (PCP)
\cite{MUNOZ}, another model
with infinitely many absorbing states.
As in the PCP, when $h=0$ all configurations
$\zeta ({\bf x})$ consistent with
$\rho_a\equiv 0$ are absorbing, and
the order parameter is coupled to a
non-order-parameter field playing
the role of an effective creation rate. 
The essential difference 
in the sandpile is that the field $\zeta$ is conserved. 
In the following we consider separately the cases 
of slow driving ($h\to 0^+$), corresponding to the SOC
sandpile, and of fixed energy: $h=0$ and $\epsilon=0$.

{\em (i) Driven sandpile:} 
The system attains its stationary state by the very slow 
addition of energy. In this limit
($h\to 0^+$), a complete time scale separation,
between toppling, on one hand, and addition and dissipation,
on the other, sets in\cite{vz,dvz,grin}.
In the stationary state, energy balance
forces a subset of the critical exponents
to take their mean-field (MF) values in any spatial dimension $d$,
as we now show.
Integrating Eq.~(\ref{cons}) over space
and averaging over the noise yields $\langle\rho_a\rangle\equiv
V^{-1}\int d^dx \langle\rho_a ({\bf x},t)\rangle=h/\epsilon$ 
\cite{vz}, which implies that the zero-field susceptibility 
$\chi\equiv\partial\rho_a/\partial h$, diverges as $\epsilon^{-\gamma}$ 
with $\gamma=1$.
Taking the functional derivative
of Eq.~(\ref{cons}) with respect to $h({\bf x'})$,
and averaging over the noise,
we obtain an equation for the static response function
$\chi_\epsilon({\bf x -x'})=
\langle\delta\rho_a({\bf x}) /\delta h({\bf x'})\rangle$: 
\begin{equation}
-D_\zeta\nabla^2\chi_\epsilon({\bf x-x'}) +
\epsilon\chi_\epsilon({\bf x-x'}) =\delta({\bf x-x'}),
\end{equation}
which yields, for large r,
$\chi_\epsilon(r)\propto r^{2-d}e^{-r/\xi} $,
where the correlation length $\xi\sim\epsilon^{-\nu}$, 
with $\nu=1/2$. These results 
depend solely upon stationarity and 
translation invariance \cite{spherical}.
Although the exponent values coincide with the MF ones,
they have not been obtained by MF arguments, and are valid beyond
MFT, as confirmed by simulations
in $2\leq d\leq6$ \cite{chessa}. 

While the remaining exponents are in principle
also determined by Eqs.~(\ref{cons}) and (\ref{active}), 
a full analysis, involving the double limit $t \to \infty$,
$h \to 0^+$ (the order cannot be interchanged),
promises to be a knotty problem.
The critical properties emerge as $\epsilon\to 0$,
which must be taken subsequent to the above limits,
since a stationary state demands $\epsilon > h$.
The upper critical dimension, however, can be  found by
power-counting analysis. The evolution of a localized 
perturbation $\rho_a$ around the slowly-driven stationary state
($h/\epsilon = 0$) is given by 
\begin{equation}
\frac{\partial \rho_a({\bf x},t) }{\partial t} =
D_a\nabla^2 \rho_a({\bf x},t) -r\rho_a({\bf x},t) 
+\mu\Delta\zeta ({\bf x},t)\rho_a({\bf x},t)  
-u\rho_a^2({\bf x},t)
+ \eta_a({\bf x},t),
\label{dyp}
\end{equation}
where $r\sim\epsilon$ and $\mu$ and $u$ are coupling terms generated 
by the elimination of  $\rho_c$ in favor of $\zeta$.
We can consider in Eq.s~(\ref{cons})and~(\ref{dyp}) the usual rescaling  
$x\to b x'$, $t\to b^zt'$ and $\rho_a\to b^{\delta_a}\rho_a'$, and the 
rescaling of the energy field $\zeta\to b^{\delta_\zeta}\zeta'$. 
The rescaled coupling constants show a MF fixed point for $r=\epsilon=0$
and $z=2$. In this case the nonlinear terms couplings scale
as $\mu\sim u\sim b^{4-d}$. Thus, nonlinear terms are irrelevant when $d>4$,
defining an upper critical dimension $d_c=4$\cite{notad_c}.
Since $r=\epsilon=0$ is the fixed point,
the dissipation rate is the (temperature-like) control parameter, 
with critical 
value $\epsilon_c=0$, emphasizing the role of conservation in 
slowly-driven sandpiles.
By studying Eq.~(\ref{dyp}) in the slowly-driven MF stationary state,
($\Delta \zeta= \rho_a=0$, $\rho_c^\infty=1/g$, and 
neglecting noise terms) we also obtain the
MFT exponents
for avalanche spreading: $D=4$, $\tau=3/2$ and $z=2$,
in agreement with earlier analysis \cite{vz}.

{\em (ii) Fixed-energy sandpile} (FES): When $h=\epsilon=0$,
the total energy
$\int d^dx \zeta ({\bf x},t)$
is conserved and plays the role
of a control parameter.
In this case,
Eq.~(\ref{cons}) reduces to 
$\partial \zeta ({\bf x},t)/\partial t=
\nabla^2 \rho_a ({\bf x},t) +\eta_\zeta({\bf x},t)$,
where $\eta_\zeta({\bf x},t)$ is a {\em conserved} noise.
Substituting the formal solution  
of this equation
into Eq.~(\ref{active}) yields
$$
\frac{\partial \rho_a({\bf x},t)}{\partial t}=
D_a\nabla^2 \rho_a({\bf x},t) 
-r({\bf x})\rho_a({\bf x},t) - b({\bf x})\rho_a^2({\bf x},t)  $$
\begin{equation}
+w\rho_a({\bf x},t) \int_0^t\nabla^2 \rho_a({\bf x},t') dt'
+\eta_a({\bf x},t),
\label{fte}
\end{equation}
where we neglect
higher-order terms, irrelevant by naive power-counting analysis. 
The coefficients $r$ and $b$ depend on position through
on the initial value of $\zeta ({\bf x})$. 
Eq. (\ref{fte}) is 
the Langevin equation of RFT, save
the non-Markovian term and the spatial variation of $r$ and $b$;
$d_c = 4$, as in RFT.
In MFT, replacing $\zeta({\bf x},0)$ by the 
spatially uniform $\zeta$, we have 
$d \rho_a/dt =-\overline{r}\rho_a 
-\overline{b}\rho_a^2$, i.e.,
the MFT of directed percolation (DP),
with critical point at $r=0$, fixing, in turn,
the critical energy density, $\zeta_c$.
Close to the critical point ($|\Delta\zeta|/\zeta_c <<1$), 
$r\sim \Delta\zeta$.  
For $d<d_c$, the critical fixed point will be renormalized to
$r=r^*$, defining a renormalized $\zeta_c$.
Above $\zeta_c$, we have an active stationary state
with $\rho_a \sim (\Delta\zeta)^{\beta}$; for $\zeta < \zeta_c$,
the system falls into an absorbing configuration in which $\rho_a=0$.

Thus the FES approach to criticality $(\zeta \to \zeta_c)$ is
fundamentally different from the driven case, ($h, \epsilon \to 0$,
followed by $h/\epsilon \to 0$).  Note that $\zeta $ is {\it lost}
as an independent parameter once $h$ and $\epsilon$ are nonzero
(Slow driving {\it pins} $\zeta$ at its critical value: if it exceeds
$\zeta_c$, activity is generated, and thereby dissipation).
The behavior {\it at} the critical point is described by our theory
with $h = \epsilon = 0$ and $\zeta = \zeta_c$.  As in other
models with infinitely many absorbing configurations, the avalanche
behavior depends intimately on the initial configuration.
It is also worth remarking that in the stationary driven case,
the dynamics can explore only a set of recurrent configurations\cite{dhar}.
The FES may instead explore transient configurations that 
could account for the different critical behavior. 

To better understand
its scaling, we simulated the FES with statistically homogeneous
initial conditions
We considered the BTW model with periodic boundary conditions
at ($\epsilon=0$,$h=0$). Initial configurations
are generated by distributing at random a fixed number $N$ of 
particles among the $L^d$ sites. This defines the strictly 
conserved energy density $\zeta=N/L^d$. Once all $N$ particles have 
been placed, active sites topple at a unit rate with a sequential
updating rule. We studied the 
transition from the active to the 
absorbing states as we varied $\zeta$.
In $d=2$, using system sizes extending up to
$L=1280$, we find
$\zeta_c = 2.125$, $\beta = 0.59(1)$,
$\nu= 0.79(4)$, and
$z=1.74(4)$ \cite{dperc}.
(Figures in parentheses denote uncertainties.)
The corresponding
DP exponents are $0.583(4)$, $0.73(2)$, and $1.76(3)$.
Simulations of the four-dimensional model yield
$\zeta_c =4.11(1)$ and $\beta = 1.00(1)$, in 
good agreement with theoretical results, which 
predict MF values in $d\geq d_c=4$ (see also Fig.2).

These results are compatible with the DP universality class, 
suggesting that the non-Markovian term is irrelevant, at least for
homogeneous initial conditions.  
Preliminary results of direct integration of Eq.(\ref{fte})
indicate DP-compatible behavior for homogeneous initial conditions;
the non-Markovian term does appear
to alter the spreading exponents, as 
in other multiple-absorbing state models \cite{pcp,MUNOZ}.
On the other hand, we find analytically that the non-Markovian
term is relevant at the RFT fixed point below $d=4$, and it has to 
be taken into account to determine the asymptotic scaling properties. 
This implies that to fully understand the effect of this term we need 
to perform a full RG perturbative expansion as well as larger 
numerical simulations.

In summary, our field theory elucidates the effect of driving on 
the critical behavior of
sandpiles through their connection with other absorbing-state
phase transitions\cite{mas}. 
Field-theoretic analysis shows that driving {\it does not} cause
such a change of critical behavior in the PCP,
as we have verified numerically.  We believe 
the crucial difference is the absence of a conservation
law in the PCP.  Indeed, a different kind of conservation
law (``local parity conservation'')
also changes the critical behavior of contact-process-like
models \cite{pc}.

Finally, we remark that our framework applies equally to 
the BTW and Manna sandpiles,
even though the latter has a stochastic toppling rule \cite{manna};
the additional noise generates no further
relevant terms.
Though it remains an open question in the context
of simulations \cite{benhur}, recent large-scale numerical studies 
and rigorous arguments support the shared universality
of the two models \cite{chessa2}. 

Our approach suggests several paths for further investigation. 
An open question concerns the critical behavior of
fixed-energy models with nonhomogeneous initial conditions.
A full renormalization group treatment of the field theory,
while challenging, should yield systematic predictions for
avalanche exponents.
Analysis of different boundary conditions should lead 
to a better understanding of the scaling anomalies
exhibited by sandpile models.

During the completion of this work, we learned of large scale
simulations of the BTW-FES model with {\em parallel}  
updating performed by P. Grassberger. The results indicate
that deviations from DP behavior persist at large system sizes,
suggesting that the non-Markovian operator in our field-theory 
could become relevant below $d_c$.

We thank A.Chessa, D.Dhar, L.Fabbian, P. Grassberger, E.Marinari and 
Y.Tu for comments and discussions. 
M.A.M., A.V. and S.Z. acknowledge partial support 
from the European Network Contract ERBFMRXCT980183. 
M.A.M. is partially supported from the TMR European 
Program ERBFMBIGT960925.

\newpage
\begin{figure}
\centerline{
       \epsfxsize=7.0cm
       \epsfbox{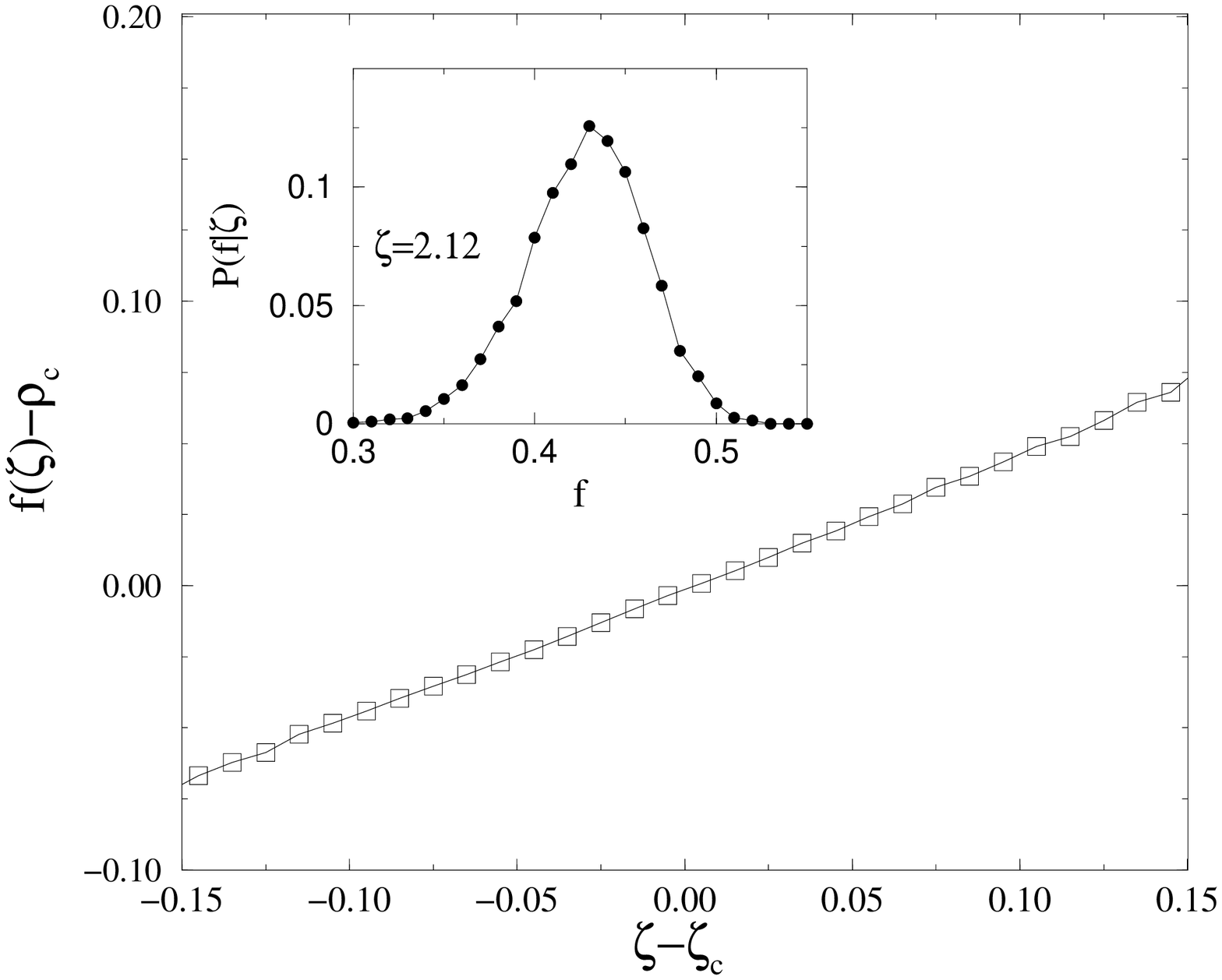}
        }
\caption{Fraction of critical nonactive sites,
$f$ (shifted by $\rho_c$)
versus
$\zeta-\zeta_c$ for the BTW in $d=2$. Inset:
conditional probability density  
at $\zeta=2.12$ for the same case.}
\label{fig:zeta}
\end{figure}

\begin{figure}[bt]
\centerline{
        \epsfxsize=7.0cm
        \epsfbox{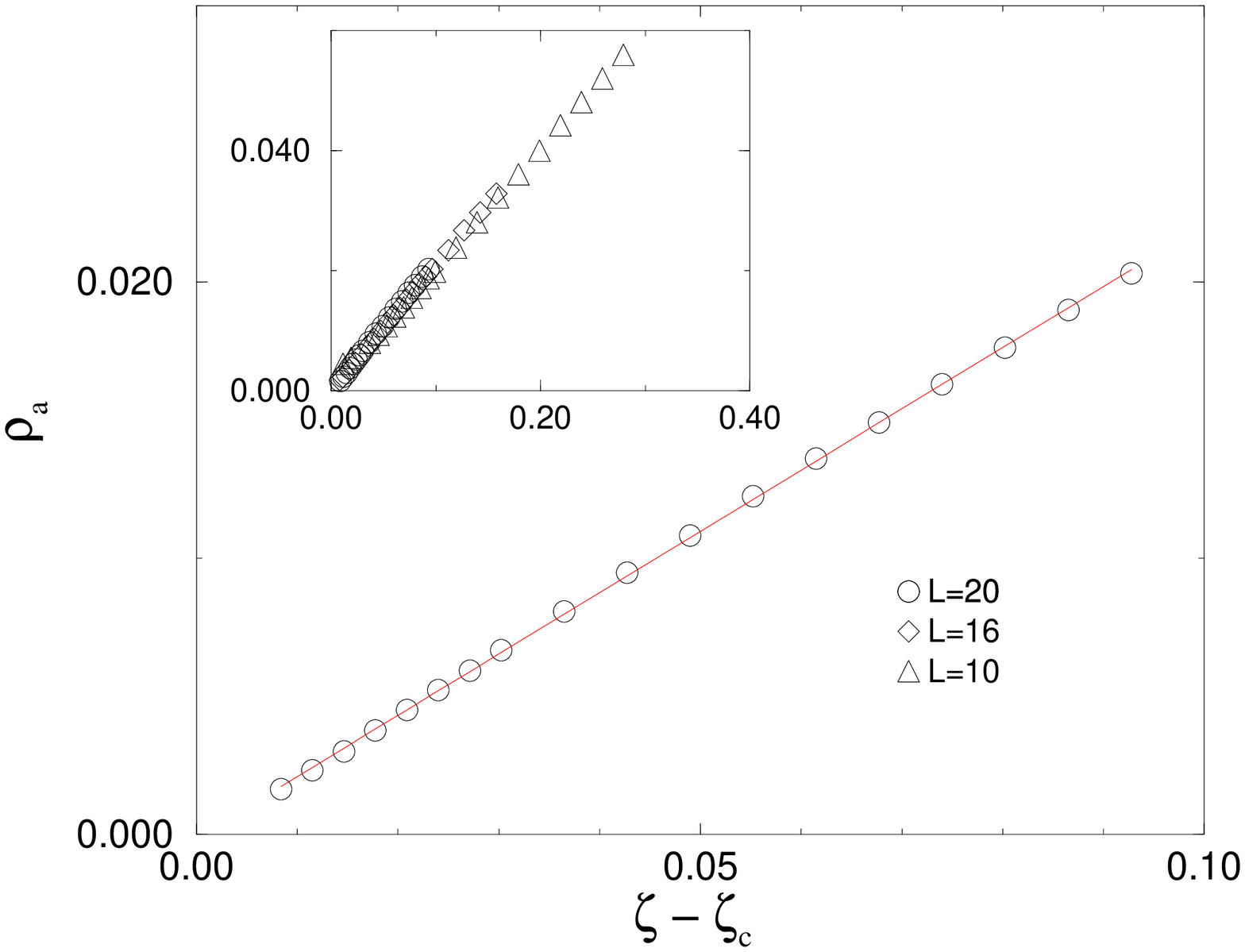}
        }
\caption{Stationary active-site density as a function of 
$\zeta-\zeta_c$ for the fixed-energy BTW model in $d=4$.
The inset shows $\rho_a$ in a larger range of $\zeta-\zeta_c$ 
and for different lattice sizes $L$. }
\label{fig:beta}
\end{figure}


\begin{thebibliography}{99}
\bibitem{top}
	$D_a$ is proportional to $\rho_c$,
	but may be treated as a constant, the neglected terms being 
	irrelevant. The negative linear 
	term represents sites that become stable after toppling. 
	Setting the toppling rate to unity effectively fixes the 
	microscopic time scale.
\bibitem{spherical} 
        For boundary dissipation,
        assuming a $d$-dimensional system of
        radius $R$, we have $\nabla^2 \langle \rho_a (r) \rangle
        = - \frac{h}{D_{\zeta}}$ for $r < R$,
        and $\langle \rho_a (r) \rangle = 0$
        at $r=R$.  This implies $\langle \rho_a (r) \rangle \propto
        hR^2 [1-(r/R)^2]$ so that the mean activity $\overline{\rho_a}
        \sim hR^2$, yielding the finite-size scaling formula
        $\chi(R) \sim R^{\gamma/\nu} $ with $\gamma/\nu = 2$.
\bibitem{notad_c}
 	It is worth remarking that the symmetry $r\propto\epsilon$ 
	has to  be explicitly enforced if we consider a single equation, 
	substituting in Eq.~(\ref{dyp}) the formal 
	integration of Eq.~(\ref{cons}). Otherwise, 
	the spurious result $d_c=6$ is obtained. From the numerical point 
	of view the value of the upper critical dimension is still a 
	debated question\cite{chessa,lubeck}.
\bibitem{dperc}
        The preliminary results reported in Ref. \cite{dvz},
        suggesting non-DP values for $\nu$ and $z$ 
        were in fact influenced by strong finite-size effects.
\end{thebibliography}
\end{document}